# An Evaluation of the Improved XP Software Development Process Model

M. R. J. Qureshi

Dept. of Computer Science,

COMSATS Institute of Information Technology Lahore Pakistan

Defence Road, Off Raiwind Road Lahore Pakistan

rjamil@ciitlahore.edu.pk

Ph # (92-42-5431602) Cell # (03334492203)

*Abstract: The concept of agile process models has attained great popularity in software (SW) development community in last few years. Agile models promote fast development. Fast development has certain drawbacks, such as weak documentation and performance for medium and large development projects. Fast development also promotes use of agile process models in small-scale projects. This paper modifies and evaluates Extreme Programming (XP) process model and proposes a novel process model based on these modifications.*

## I. Introduction

Agile process models stress on agility for software development. Agility signifies responding to changes quickly and efficiently. Possible changes required in software projects are in budget, schedule, resources, technology, requirements and team. These are "reacting" changes on which agile models stress. They are called agile golden principles that are defined in agile alliance meeting conducted in 2001 [1].

The aim of agile principles is to have adaptive software development only for simple and small size software projects. There is no indication to adapt process models according to nature of the projects. Analyst has to select traditional software process models if the software is average or complex in nature, such as Spiral and Rational Unified Process (RUP). Section 2 proposes an enhanced XP process model for agile and traditional software development. Section 3 describes how case study is adapted to validate enhanced XP model on small, medium and large projects. Section 4 shows empirical evaluation based on data gathered from thirty eight professionals from seven software development organizations.

## II. An Enhanced XP Process Model for Agile Development

The proposed process model is a modified approach of XP model, which is most widely practiced model among agile models. The main phases of XP model are planning, design, coding and testing [2]. The main phases of adaptive model are Project Planning, Analysis and Risk Management, Design & Development and Testing.

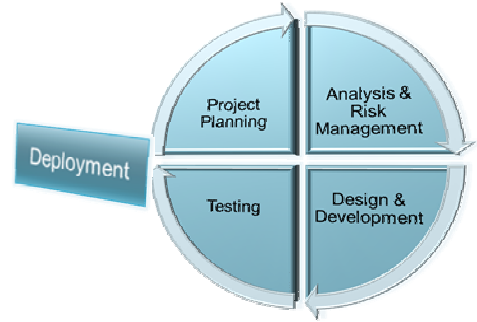

Figure 1 An Enhanced XP Process Model

The focus of the proposed process model is the implementation on medium and large scale projects which keep on evolving due to changes in customer requirements. Project specification or proposal document is prepared during the '*Project Planning*' phase by communicating to the customer. Project specification or proposal document is composed of feasibility report that is created to prepare a cost benefits analysis (CBA) sheet. Feasibility report is composed of economic, technical and operational feasibilities. Organizational feasibility is also prepared based on client request or if project demands. CBA sheet helps to estimate whether SW project is feasible for the customer or not. Project team members are also selected during planning phase. Project team size depends on the size and schedule of project.

'*Analysis & Risk Management*' phase is only started if a customer approves the proposal. Analysis phase improves quality of software through proper documentation. This is the phase in which an analyst gathers detailed requirements. Software requirement specification (SRS) document is produced at the end of analysis phase. Main contents of SRS are: summary of requirements, requirements modeling, data modeling and risk management plan. Requirements could either be structured or object oriented (OO). Entity relationship diagram (ERD) is drawn in case of structured development and object diagram is created in case of object oriented development.

*Design and development* phases of XP model are merged to incorporate agility in the proposed process model. The enhanced XP process model uses the prototype approach to verify the design and requirements. Software is developed incrementally as customer approves prototypes.

Test cases are prepared for each increment at the start of *Testing* phase. Each module is tested on unit basis. Integration test is then performed to check integration among modules. System test is the next phase to validate the whole increment as one unit. Acceptance testing is the last test to verify increment from the customer. Tested increment is maintained and deployed. The proposed process model is cyclic and evolutionary till whole software is developed.

## III. Case Study: Strategy to Implement Enhanced XP Process Model

A strategy is recommended in this paper to schedule and manage medium and large size projects. This strategy is based on an evaluation study of a software organization. The proposed process model is made adapted according to SW organization and according to nature of project. Each phase of the process model is described as follows with activities performed in this project. The success of this case study reflects that there is a possibility to make XP adaptive, successful and practical in SW industry environment for medium and large size projects.

**Case study**

It was conducted with a software organization which has successfully implemented adaptive process model on a project for two leading companies of USA dealing in property estate business. Their market share in property estate business is around 70-80%. This business estate spans whole USA.

**Project Planning Phase**

Two Product specialists met with the client companies on their request. The request was to develop an information system for property estate business. They gathered basic requirements which were called as initial user stories. Economic, technical, operational and schedule feasibilities were prepared based on these stories. Customers approved the feasibility report. Six years were estimated to complete this Project. Project has been broken down into several milestones. Each milestone is a complete project itself ranging from one and half month, four months and six months. Product specialists and Project Manager decided that each project will be divided into builds.

Each build is an increment based on number of stories completed in it. Two week time has been decided to design, develop, test and deploy a build following agile principles. Each team will be working on a particular build. That means four builds will be designed, developed, tested and deployed at the same time.

Four teams are dedicated for this project. Each team is working on different domain area. Team size for overall project is as follows.

- Two Product Specialists (Each product specialist handles to two teams)
- One Project Manager (shared among four teams)
- One Technical Architect in each team
- Two Database developers in each team (One for One and half month project)
- Two-Four Programmers in each team ( Two for One and half month)
- One Configuration Manager shared to four teams
- One Quality Assurance team leader shared among four teams
- One-Two Quality Assurance engineer in each team
- One Documentation incharge in each team

**Analysis & Risk Management Phase**

Analysts were called product specialist in this software organization. Two product specialists were deputed to work on this project. Each product specialist in this project has four major roles.

- Coordinates between customer and SW House.
- Gathers and writes all user stories from customer.
- Writes initial use cases.
- Approve functionalities of each increment or build.

Three months were spent to gather complete user stories during analysis phase from both the companies dealing in estate business. Client companies were requested to prioritize user stories and to assign a scale value to each used story like XP. Risks regarding these user stories were estimated. User stories with high scale value and high risk values are moved up in the index to be implemented initially. User stories are aggregated according to the domain area.

**Planning Poker Technique**

Project Manager (PM) conducted a meeting with each team to implement planning poker technique. All team members were present during the meeting. PM selected first top 10 user stories about a particular domain for a meeting with first team to design, develop, test and deploy for the first build. He selected 10 user stories (according to domain area) to conduct similar meetings with remaining 3 teams. PM conducted meeting with all teams differently after every two weeks to implement planning poker technique on next user stories according to index values. Scenario of a meeting is as follows.

Odd numbers of cards were distributed among all team members. PM instructed team members to assign story points to each user story. Card values ranged from 1 to 21. PM started meeting with the user story # 1 and briefly described it with in 2 to 3 minutes. He asked team members to assign a story point value which is manageable in order to design, develop, test and deploy this user story. One development member raised card with a point value one. Technical architect raised card with point value of 13. Deployment member raised card with a point value 21. Quality assurance engineer raised card with point value 9. PM asked each member to provide a logical reason that why he/she assigned this story point value to user story in discussion. PM asked them to repeat the whole exercise and assigned a story point value again. This exercise was continued until all team members agreed on a manageable story point value such as 7. In this way all 10 user stories were explained and assigned a point value.

Card point technique meeting was conducted after every two weeks. PM measured progress of the project after 4 weeks on completion of 8 builds. The story point values were ranging from 81 to 121 for these builds. PM decided that achievable range for four teams was 81 story points for rest of the builds to design, develop, test and deploy. He stopped the story point meeting when user stories for a build scored 81 story points. Remaining user stories were adjusted in the next build.

**Cases to Adjust User Stories**

- User story was adjusted in the next build in a case if 81card points were not achieved due to any problem during designing or coding or testing or change in customer requirements.
- User stories were also adjusted in the next build when customer asked to change a build already deployed. Customer asked to bring these changes due to change in his requirements. These user stories were considered as new user stories and were evaluated again using planning poker technique to design, develop, test and deploy.

Initial use cases were developed for the user stories selected for every build during analysis phase. Function specification was composed of indexed user stories and initial use cases. Functional specification kept on changing during life of the project based on customer requests.

**Design and Development Phase**

Design and development was performed in parallel according to the adaptive process model in contrast to XP process model where these were separate phases.

Quality assurance engineer designed detailed use cases for the user stories to be implemented after first two weeks. Entity relationship diagram (ERD) and object diagrams were made. Normalization was made based on object diagram and ERD. Database was physically designed based on normalization. All these tasks were performed by database developers. Technical architect designed the architecture of the user stories for a build. He was also responsible for deploying a particular build. He was also responsible for running a build on local area network (LAN) or on Wide area network (WAN) or dot net remote services required. User interfaces were also defined for the user stories to be developed for a build. Each user interface had a complete description regarding functions, inputs, outputs and which user story/ies was/were going to be accomplished. Technical specification was completed at this level. Technical architect prepared technical specification. Technical specification composed of detailed use cases, ERD, normalization, database construction, architecture, deployment details and user interface specifications (user interface description).

After particular user interface specification was completed it was assigned to programmer for coding. Programmer prepared initial test cases for integration and system tests. Test cases for unit testing were called as unit tests. Unit tests were prepared by programmers completely. Quality assurance engineers were informed to prepare complete test cases about an interface to conduct system and acceptance tests. All test cases were provided to programmers within 2 to 3 days. The user interface specification was completed by the time programmer wrote code for an interface already delivered. This process was repeated for all the remaining user stories to be completed in every build.

**Testing Phase**

Programmers conducted unit, integration and system tests for every build using case tools. NUnit [3] and FitNesse [4] SW were used to conduct unit, integration and system tests. NUnit SW was used to conduct unit testing and FitNesse SW was used to conduct integration and system tests. NUnit and FitNesse SW both provided graphical user interface (GUI). NCover [5] SW is a case tool that uses NUnit software to check how much code is covered during testing for a particular business object. In this test each line should be tested.
Programmers first run NUnit software to conduct unit tests. Programmers then run NCover that used NUnit to check the code coverage. Programmers conducted integration and system tests using FitNesse software. Programmers integrated user stories into visual source safe as these are completed. Configuration Manager (CM) compiled all user stories completed in a build to create executable (EXE) file. CM asked to programmers to fix code if it did break and EXE file was not created. Quality assurance manager asked to quality assurance engineers to test code if did not break and EXE file is created. Quality assurance engineer tested entire code again for a build using NUnit, Fitnesse and NCover SW. Build was ready for workable demo for the product specialist if it was approved by quality assurance engineers. CM labeled builds after approval of product specialist. Eighteen hundred and ninety seven builds were completed within 22 months and four milestones were achieved and deployed for this project.

# IV. Evaluation of the Enhanced XP Process Model

A survey involving seven software development organizations was conducted to evaluate the enhanced XP process model. A list of software houses was taken from Pakistan Software Export Board. Software houses were selected randomly to gather the data which were dealing globally. Team sizes of software houses were twenty to one thousand. Questionnaire technique was used to gather the data. Thirty eight professionals were selected to fill the questionnaire forms. The people who filled the forms had more than six years experience in software development. Lacquered scale was ranging from 1 to 5 to gather the data against the questionnaires. Questionnaire was divided into three main sections. Each section was consisted of questions. Core objective of each section was to adapt XP process model for small, medium and large projects. The three sections were addressing following main areas.

- Suitability of changes into the XP process model for medium and large projects.
- Affect of incorporation of 'Analysis and Risk Management (ARM) phase into XP process model.
- Affect of merging design and development (DD) phases of XP process model.

**Section 1- Suitability of Changes in XP Model for the Development of Medium and Large Projects**

Table 1 has been created on the basis of evaluations. The parameters evaluated in Table 1 were as follows:

A. Rank 'Project Planning' phase suitability for adaptive process model.
B. Does analysis and Risk Management phase affect in development of average and complex software projects?
C. Does 'Testing' phase of XP process model suite to test medium and large size software projects?

| Weight | % of 1 | % of 2 | % of 3 | % of 4 | % of 5 |
|---|---|---|---|---|---|
| Parameter | | | | | |
| A | | 2.6 | 28.9 | 52.6 | 15.8 |
| B | | 7.9 | 31.6 | 34.2 | 26.3 |
| C | 2.6 | 10.5 | 26.3 | 34.2 | 26.3 |

**Table 1-Suitability of Changes into XP model for Medium and Large Projects**

Table 1 shows affect of changes in the XP process model. It shows that most of SW developers are of the view that changes in the XP process model has high affect in the development of medium and large projects.

**Section 2- Affect of Incorporation of Analysis and Risk Management Phase into the XP Process Model**

Table 2 has been created on the basis of evaluation of ARM phase. Parameters evaluated in 'Analysis and Risk Management' phase were as follows.

A- Addition of Analysis and Risk Management phase caters the potential risks in development of medium and large projects.
B- Analysis and Risk Management phase incorporation into XP makes it adaptive for all types of projects.
C- Analysis and Risk Management phase facilitates to gather complete requirements.
D- Analysis and Risk Management phase helps in better design.
E- Analysis and Risk Management phase helps to verify/define better interfaces.
F- Analysis and Risk Management phase improves factor of reuse.
G- Analysis and Risk Management phase helps to have better documentation.
H- Analysis and Risk Management phase improves quality of SW.
I- Analysis and Risk Management phase affect on efficiency of process model
J- Analysis and Risk Management phase affect on efficient development

| Weight | % of 1 | % of 2 | % of 3 | % of 4 | % of 5 |
|---|---|---|---|---|---|

| Parameters | | | | | |
|---|---|---|---|---|---|
| A | | 2 | 31 | 47 | 21 |
| B | 2.6 | | 31.6 | 50 | 15.8 |
| C | 2.6 | 7.9 | 21.1 | 50 | 18.4 |
| D | | 2 | 5 | 34 | 57 |
| E | 2.6 | 5.3 | 21.1 | 47.4 | 23.5 |
| F | | 2.6 | 23.7 | 42.1 | 31.6 |
| G | | | 28.9 | 52.6 | 18.4 |
| H | | 2.6 | 10.5 | 55.3 | 31.6 |
| I | | 7 | 21 | 50 | 21 |
| J | | 5.3 | 15.8 | 31.6 | 47.4 |

**Table 2- Affects of Analysis and Risk Management Phase**

The results show that most of respondents of the view that 'Analysis and Risk Management' phase is very important for development of medium and large projects.

**Section 3- Affect of Merging Design and Development (DD) Phases**

The parameters evaluated in Table 4 were as follows.

- A- Merging (design and development phases) incorporates agility into XP process model.
- B- Merging (design and development phases) improves overall efficiency of XP process model.
- C- Merging (design and development phases) improves efficiency of SW development.

| Weight | % of 1 | % of 2 | % of 3 | % of 4 | % of 5 |
|---|---|---|---|---|---|
| Parameters | | | | | |
| A | | 7.9 | 34.2 | 44.7 | 13.2 |
| B | | 13.2 | 31.6 | 44.7 | 10.5 |
| C | | 2.6 | 36.8 | 36.8 | 23.7 |

**Table 4- Evaluation of Merging Design & Development**

It can be concluded from Table 3 that merging design and development phases of XP model improves agility and efficiency of process model. The merging also facilitates SW development.

Frequency of feedback from Table 1 to 3 supports the view that XP model can be adapted for medium and large size projects.

## V. Conclusion

This paper supports practice of agile software development by proposing a process model which is adapted according to the requirements of the software project. The enhanced XP process model is better than XP because it eliminates the limitations of development of reusable components, large development teams, documentation, medium and large software projects.